\documentclass[twocolumn]{article}
\usepackage[left=2.5cm,right=2.5cm,top=2.5cm,bottom=2.5cm]{geometry}
\usepackage{moreverb,url}
\usepackage{makecell}
\usepackage{graphicx}
\usepackage{subcaption}
\usepackage[colorlinks,bookmarksopen,bookmarksnumbered,citecolor=red,urlcolor=red]{hyperref}
\usepackage{booktabs}

\begin{document}


\title{Egocentric Video: A New Tool for Capturing Hand Use of Individuals with Spinal Cord Injury at Home}

%

\author{Jirapat Likitlersuang\textsuperscript{1, 2}, Elizabeth R. Sumitro\textsuperscript{1, 2}, Tianshi Cao\textsuperscript{2}, \\ Ryan J. Vis\'ee\textsuperscript{1, 2}, Sukhvinder Kalsi-Ryan\textsuperscript{2, 3}, and Jos\'e Zariffa\textsuperscript{1, 2}}

\date{\textsuperscript{1} Institute of Biomaterials \& Biomedical Engineering, University of Toronto, Canada\\ \textsuperscript{2} Toronto Rehabilitation Institute, University Health Network, Canada \\ \textsuperscript{3} Department of Physical Therapy, University of Toronto, Canada \\ \today}




\maketitle

\begin{abstract}
\textbf{Background:} Current upper extremity outcome measures for persons with cervical spinal cord injury (cSCI) lack the ability to directly collect quantitative information in home and community environments. A wearable first-person (egocentric) camera system is presented that aims to monitor functional hand use outside of clinical settings.\\  
\textbf{Methods:} The system is based on computer vision algorithms that detect the hand, segment the hand outline, distinguish the user’s left or right hand, and detect functional interactions of the hand with objects during activities of daily living. The algorithm was evaluated using egocentric video recordings from 9 participants with cSCI, obtained in a home simulation laboratory. The system produces a binary hand-object interaction decision for each video frame, based on features reflecting motion cues of the hand, hand shape and colour characteristics of the scene. \\ 
\textbf{Results:} The output from the algorithm was compared with a manual labelling of the video, yielding F1-scores of $0.74 \pm 0.15$ for the left hand and $0.73 \pm 0.15$ for the right hand. From the resulting frame-by-frame binary data, functional hand use measures were extracted: the amount of total interaction as a percentage of testing time, the average duration of interactions in seconds, and the number of interactions per hour. Moderate and significant correlations were found when comparing these output measures to the results of the manual labelling, with $\rho = 0.40, 0.54$ and $0.55$ respectively. \\
\textbf{Conclusions:} These results demonstrate the potential of a wearable egocentric camera for capturing quantitative measures of hand use at home.  
\end{abstract}

{\scriptsize
\noindent
\textbf{Corresponding author:} Jos\'e Zariffa (jose.zariffa@utoronto.ca)
}

\section{Introduction}
Upper extremity (UE) impairment can severely limit individuals' ability to perform activities of daily living (ADLs). The recovery of hand function is consequently of great importance to individuals with cervical spinal cord injuries (SCI) \cite{RefWorks:138}.

In order to assess new interventions and improve upon currently existing rehabilitation approaches, outcome measures that can accurately quantify hand function in a natural context are needed. The majority of currently available assessment tools measure impairment or functional limitation and rely on direct observation by a clinician or investigator \cite{KalsiRyan2012}. On the other hand, limited tools are available to describe how individuals with SCI use their hands in their usual environment or community, restricting our understanding of how changes in UE function impact activity and participation. Often measures that describe function in the community are restricted to potentially biased self-report to try to estimate independence in ADLs \cite{linacre1994, Heinemann1993, Catz1997, Itzkovich2007}. 

In an attempt to gauge hand function at home and in the community, multiple studies have explored the use of wearable sensors, particularly accelerometers or inertial measurement units (IMUs). While these devices are small and easily worn, they lack the resolution to capture the complexity of functional hand use. These wearable sensors are limited to capturing arm movements. In studies with hemiparetic stroke survivors, accelerometry has most typically been used to examine the ratio of activity between the impaired and unimpaired arms \cite{RefWorks:148, RefWorks:149}. Recently in SCI, accelerometers have been used to measure wheeling movements and to assess the laterality of the injuries \cite{Brogioli2016, Brogioli2016_2}. Although these studies demonstrated a relationship between accelerometry measures and independence, this approach is not able to reveal direct information about how the hand is used in functional tasks \cite{Brogioli2016_2}. A recent study in stroke survivors found that improvements in motor function according to clinic-based outcome measures (capacity, as defined in Marino’s modification of the International classification of functioning, disability and health (ICF) model \cite{RefWorks:139}) do not necessarily translate into increased limb use in the community (performance), as measured by accelerometry \cite{RefWorks:147}. These findings point to the need for novel outcome measures that can directly measure performance and better describe the impact of new interventions on the daily lives of individuals with SCI.

A system based on a wearable camera has the potential to overcome the limitations of existing outcome measures for UE function. First-person (egocentric) cameras record the user’s point of view. Unlike a wrist-worn accelerometer that can only capture arm movement information, an egocentric video provides detailed information on hand posture and movements, as well as on the object or environment that the hand is interacting with. Multiple studies have explored the use of computer vision techniques to extract information about the hand in egocentric videos, though typically not in the context of rehabilitation. Key problems to be solved include hand detection (locating the hand in the image) as well as segmentation (separating the outline of the hand from the background of the image) \cite{RefWorks:154,RefWorks:155,RefWorks:156,RefWorks:151,RefWorks:163,RefWorks:168,RefWorks:169, Zariffa2013, Khan2018, Bambach2015}. Beyond hand detection and segmentation, there have also been attempts to use egocentric videos for activity recognition and object detection in ADLs \cite{RefWorks:162,RefWorks:161,RefWorks:160,RefWorks:166,RefWorks:165,RefWorks:170}. However, generalizability can be a challenge in such systems, given the large variety of activities and objects found in the community.

In our previous work, we proposed to detect interactions of the hand with objects using egocentric videos. This binary classification (interaction or no interaction) is intended to form the basis for novel outcome measures to describe hand function in the community. We have demonstrated, in the able-bodied population, the possibility of a hand-object interaction detection system, where a system can detect and log whether or not the hand is manipulating an object for a functional purpose. An interaction between an object and the hand is only considered to happen when the hand manipulates the object for a functional purpose; for example, resting a hand on the object would not constitute an interaction \cite{RefWorks:170}. The present work expands the development of the hand-object interaction detection system by describing a novel complete algorithmic pipeline and evaluating for the first time its application to individuals with SCI.

\section{Methods}
\subsection{Dataset and Participants}
A dataset from participants with cervical SCI was created, the Adaptive Neurorehabilitation Systems Laboratory dataset of participants with SCI ("ANS SCI"). The ANS SCI dataset consists of egocentric video recordings reflective of ADLs obtained using a commercially available egocentric camera (GoPro Hero4, San Mateo, California, USA) worn by the participant overhead via a head strap. The video was recorded at 1080p resolution at 30 frames per second. However, a reduced resolution of 480p was used for analysis in order to reduce computation time, given that higher resolutions are not necessary for our application. The data collection was performed in a home simulation laboratory. Specifically, this study involved recording from 17 participants with SCI, performing different common interactive ADL tasks identified by the American Occupational Therapy Association (AOTA) as important (for example, personal care, eating, and leisure activities)  \cite{RefWorks:86}. EEach participant performed a total of approximately 38 ADL tasks in several environment (kitchen, living room, bedroom, bathroom). Participants were also asked to perform non-interactive tasks, which involved hand at rest and hand being waved in the air without any interaction with an object.

\begin{table*}
\small\sf\centering
\caption{Participant Demographics and Injury Characteristics. \label{T1}}
\label{table:T1}
\begin{tabular}{cccccccc}
\toprule
Participant & Age(Years) & Sex & Level of Injury & AIS grade & \makecell{Tramatic(T)/\\Non-tramatic (NT)} & \makecell{Time since \\ injury (Years)} & \makecell{Upper Extremity\\Motor Score (UEMS)}\\
\midrule
\texttt{1} & 63 & Male & C5-C6 & A* & T & 8 & 15\\
\texttt{2} & 58  & Male & C3-C5 & D & T & 1 & 24\\
\texttt{3} & 59  & Male & C2-C6 & D & T & 1 & 20\\
\texttt{4} & 55  & Male & C7-T1 & C/D*  & T & 4 & 18\\
\texttt{5} & 56  & Male & C2-C7 & D & T & 2 & 19\\
\texttt{6} & 56   & Male & C5-C6 & D & T & 2 & 16\\
\texttt{7} & 20  & Male & C5 & B & T & 4 & 9\\
\texttt{8} & 58  & Male & C5 & C/D* & T & 32 & 13\\
\texttt{9} & 44  & Female & C6-C7 & A& T & 20 & 20\\
\texttt{10} & 51   & Male & C4-C6 & D & T & 1 & 22\\
\texttt{11} & 34  & Male & C5-C6 & C & T & 5 & 21\\
\texttt{12} & 40  & Female & C2-T1 & D & NT & 2 & 20\\
\texttt{13} & 70  & Male & C4-C6 & C & T & 1 & 24\\
\texttt{14} & 42 & Male & C4-C6 & B & T & 0.4 & 16\\
\texttt{15} & 52 & Male & C1-C6 & D & NT & 0.3 & 23\\
\texttt{16} & 44 & Male & C4-C5 & B* & T & 21 & 21\\
\texttt{17} & 41 & Male & C6-C7 & A* & T & 20 & 14\\
\multicolumn{8}{l}{*These AIS grades are based on self-report}\\
\bottomrule
\end{tabular}
\end{table*}

\subsection{Algorithmic framework}
\begin{figure*}
\includegraphics[width=\textwidth]{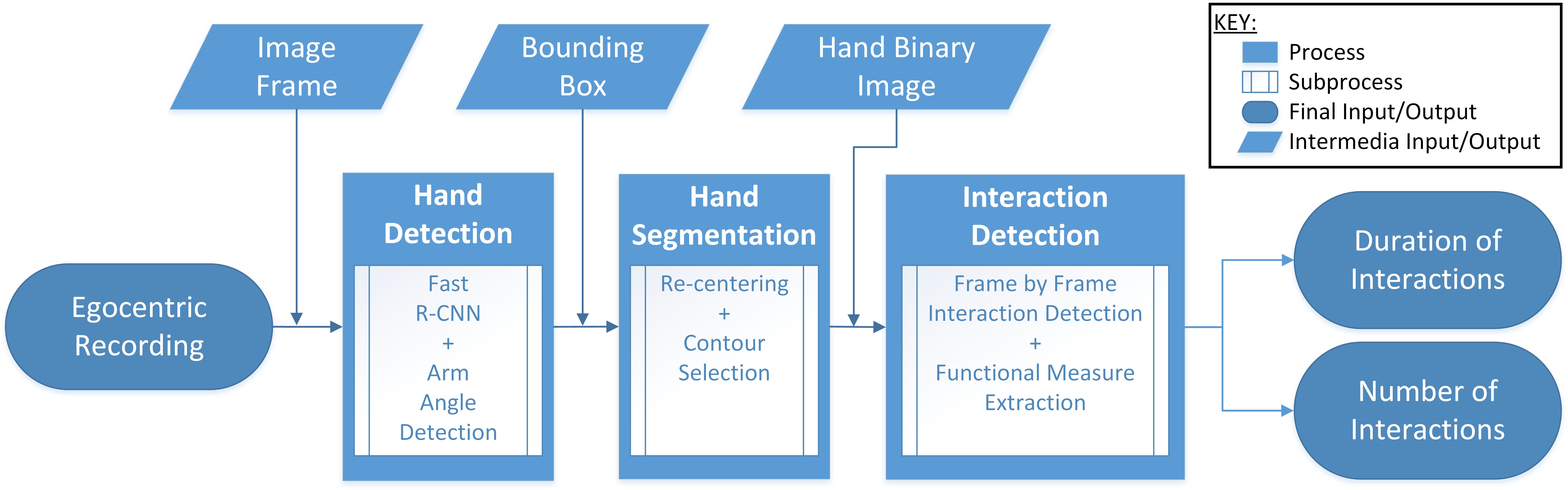} 
\caption{A simplified flowchart of the algorithmic framework showing the developed sequential preprocessing steps as well as input and output format for each step} 
\label{fig:F1}
\end{figure*}

In order to capture the hand-object interactions, the framework developed consisted of three processing steps. First, the hand location was determined in the form of a bounding box. Next, the bounding box was processed for hand segmentation, where the pixels of the hand were separated from the non-hand pixels (i.e. the background). With the hand being located and segmented, image features including hand motion, hand shape and colour distribution were extracted for the classification of hand-object interaction. The flowchart in Fig. \ref{fig:F1} summarizes the algorithmic framework. Note that our previous work in \cite{RefWorks:170} focused on the interaction detection step, whereas in the present study the hand detection and segmentation steps have been developed and included, allowing us to evaluate the complete pipeline from raw video to interaction metrics.

\begin{figure*}
	\centering
	\begin{subfigure}[b]{\textwidth}
	\textbf{(a)}
	\includegraphics[width=\textwidth]{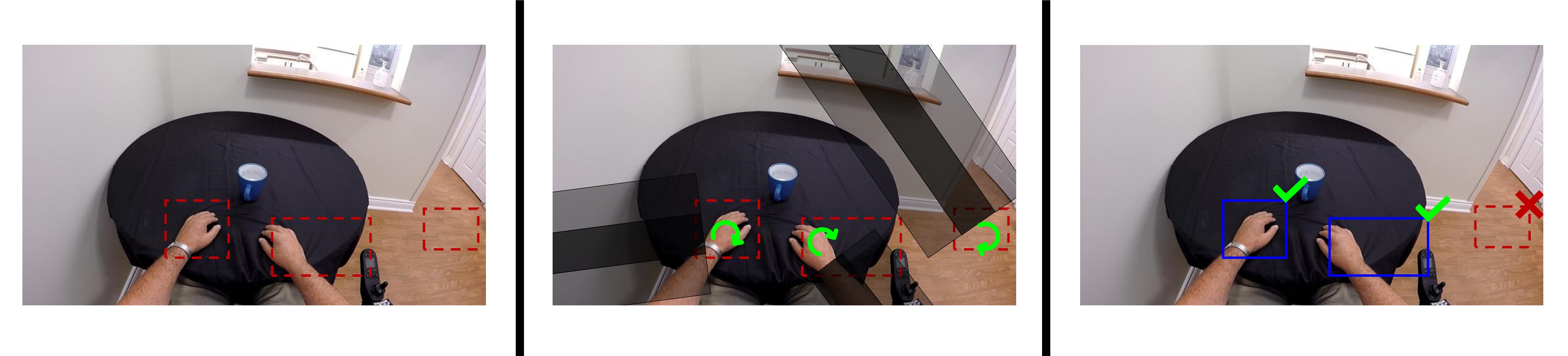}
	\label{fig:F2a}
	\end{subfigure}
	\begin{subfigure}[b]{\textwidth}
	\textbf{(b)}
	\includegraphics[width=\textwidth]{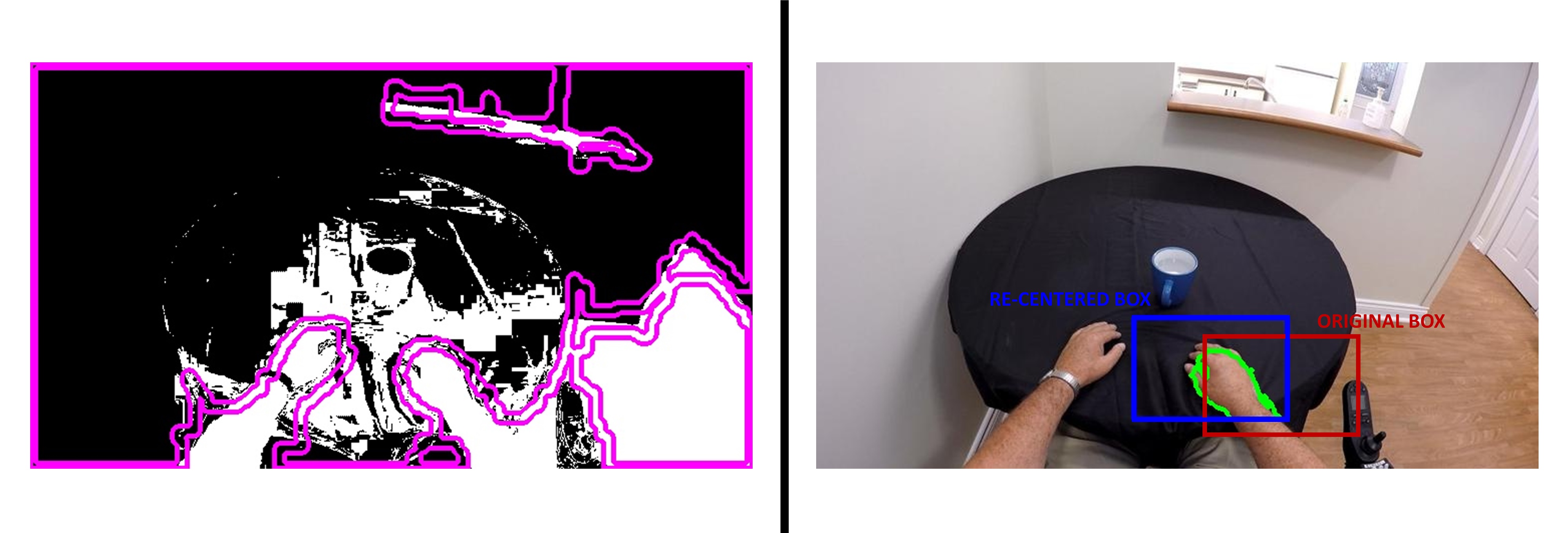}
	\label{fig:F2b}
	\end{subfigure}
	\begin{subfigure}[b]{\textwidth}
	\textbf{(c)}
	\includegraphics[width=\textwidth]{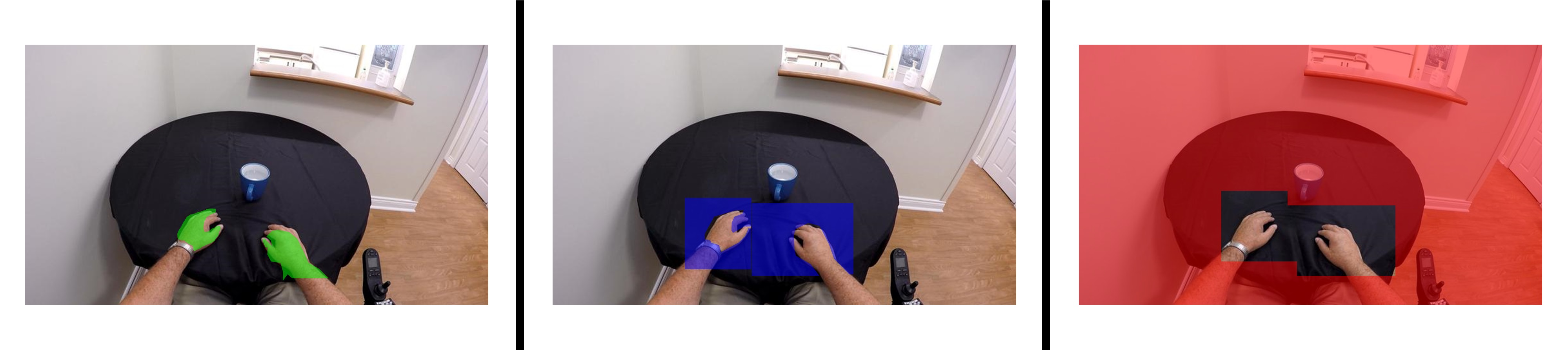}
	\label{fig:F2c}
	\end{subfigure}
\caption{Example frames describing the methodology in each of the processing steps. (a) Hand detection step, where the left image is the output bounding box of the hand from the R-CNN, the centre image is the Haar-like feature rotating around the bounding box centroid, and the right image is the final detection output. (b) Hand segmentation step, where the left image is the hand contour identification generated by combining skin colour information (in black and white) with edge detection of hand contours (in purple), and the right image shows the re-centering and selection of the final hand contour. (c) Regions involved in the interaction detection step, where the left image is the hand region, the centre image is the boxed neighbourhood of the hand, and the right image is the background region}
\label{fig:F2}
\end{figure*}

\subsubsection{Hand Detection:}

The detection of the hand was accomplished using a convolutional neural network (CNN). Specifically, the faster regional CNN ("faster R-CNN") architecture \cite{Ren2015faster} was selected as it was shown to improve detection speed over traditional R-CNN. The CNN outputs the coordinates of the box surrounding the detected hand (Fig. \ref{fig:F2}a). The CNN was trained on the even frames of eight participants in our dataset (participants \# 10-17) resulting in 33,256 frames manually labelled with hand bounding boxes.

The output bounding box from the CNN was further processed to help eliminate false positives and to extract arm angle information, using a rotating Haar-like feature.  This feature consists of three adjacent parallel rectangles extending from the centre and sides of the bounding box to the border of the image; this was inspired by the Haar-like feature used in real-time face detection \cite{Viola2004}. This selected centre rectangular region was compared to the two parallel regions on either side, by taking the average of the difference between the coefficient of variation in the centre rectangle and in each of the two side rectangles; this difference is expected to be greatest when the center rectangle is aligned with the arm. This Haar-like feature was rotated 360 degrees around the centroid of the hand bounding box. The output from the rotating feature, computed in 1-degree step was then summed over bins of 5 degrees to obtain a feature vector with 72 values. This vector was used as input to a random forest classifier with binary output, to confirm whether the bounding box truly included a hand (Fig. \ref{fig:F2}a). The training of this classifier was performed on the odd frames of eight participants (i.e. participants \# 10-17, the same ones used to train the CNN), which again consisted of 33,256 frames labelled with bounding boxes.

Additionally, given the egocentric viewpoint, the arm angle was used to determine which hand was being used (i.e. user's left or right hand, as well as a hand belonging to another individual). This was established by summing the values in the Haar-like feature vector in the top two quadrants (Quadrants I+II, 0-180 degrees), bottom right quadrant (Quadrant III, 180-270 degrees), or bottom left quadrant (Quadrant IV, 270-360 degrees). The hand was determined to be the left hand, right hand or other person's hand if the quadrant with the highest sum was IV, III or I+II, respectively.

\subsubsection{Hand Segmentation:}

The output of the hand detection stage was processed to segment the hand (i.e. identify pixels belonging to the hand). The segmentation process consisted of the following steps:

1. Identify candidate hand contours. This step combined colour and edge information. For colour, the RGB image was back-projected using a histogram obtained from a generic mixture-of-Gaussians skin colour model \cite{Jones2002}. The back-projected image was thresholded at 0.75 of the maximum value (this value was selected empirically and may change depending on the camera system). To obtain edge information, a Structured Forests edge detection \cite{Dollar2013} method was used, which was specifically trained on hand images to preferentially identify hand edges. For the purposes of training this model, a publicly available dataset \cite{Betancourt2014} with pixel-level hand annotations \cite{Cartas2017} was used. The output of the edge detection was thresholded at 0.05 of the maximum value (again selected empirically), and morphological operations (dilation followed by erosion) were used to remove small gaps in the contours detected. Lastly, the edge information was used to improve the delineation of the hands in the colour-based segmentation (Fig. \ref{fig:F2}b).

2. Re-centre the bounding box and select the final hand contour. The bounding box from the hand detection step was applied to the image from step 1. Within this box, we sought to identify the contour most likely to be the hand and re-centre the box around it to minimize the occurrence of truncated hands.  The determination of this final hand contour was based on shape, again using the information from the edge detection. From the list of contours obtained in the previous step from the combined colour and edge information, we selected the one that had the highest overlap with the edge image, filled in using dilation.  Prior to this determination, any contour whose area was less than 2\% or more than 75\% of the bounding box area was eliminated; similarly, contours with arc lengths between 90\% and 110\% of the bounding box perimeter were removed. Once the hand contour was determined, a new box and associated centroid was selected from the mean of the hand contour's centroid and top pixel. This step promotes maximum coverage of the hand and not the arm, which is often located below the hand in an egocentric view.

\subsubsection{Interaction Detection:}

The interaction detection is built on our previous work \cite{RefWorks:170}, with additional colour features included. Furthermore, the interaction detection for this study also supported multiple hands, with the user's left or right hand as well as the hands of other individuals being identified using the arm angle obtained via the hand detection step. Three categories of features were extracted in the interaction detection:

1. Object Motion. Motion features assume that an object being held in the hand will be moving with a similar direction and speed as the hand. Conversely, an item in the frame that is not being interacted with is more likely to have a motion similar to that of the background. To capture this distinction, a dense optical flow map \cite{Farneback2003} was separated into three regions: the segmented hand, the bounding box around the hand, and the background (Fig. \ref{fig:F2}c). Note that the bounding box size and location are equivalent to the segmentation step after re-centering. The dense optical flow from each region was summarized into respective histograms of magnitude and direction, each with 15 normalized bins. The final feature consisted of two vectors: the subtraction of the histograms of the bounding box near the hand from those of the hand, and the subtraction of the histograms of the bounding box near the hand from those of the background. 

2. Hand shape. Certain characteristics of hand shape are indicative of hand activity (i.e. grip type). The hand shape was represented using histograms of oriented gradients (HOG), implemented as in our previous work \cite{RefWorks:170}. The HOG features were extracted from the same bounding box used in the motion feature analysis above. The selected image regions were then resized to 10\% of the frame height and 15\% of the frame width to guarantee identical dimensions before principal component analysis (PCA) was applied. The HOG feature vector dimension was reduced from 960 to 60 to keep the dimensions identical to that of the motion cues feature. 

3. Object Colour. The objects near the hand may have a different colour than the background and the hand. The closer the objects to the hand, the greater the likelihood of interaction. HSV colour histograms were extracted from the same three regions as in the motion feature analysis described earlier. The two comparison scores are extracted based on the Bhattacharyya distance between the histograms of the bounding box and the hand, as well as those of the bounding box and the background. 

Finally, the combined features were input to a random forest binary classifier  (see “System Evaluation” below). The random forest used 150 trees, following the work in \cite{RefWorks:170}.

The classifier was trained using data manually labelled by a human observer, where each frame was either classified as interaction or no interaction. An interaction between the object and the hand was only considered to happen when the hand manipulates the object for a functional purpose. For example, resting a hand on the object or moving a hand through space would not constitute an interaction. Labelling was performed on a frame-by-frame basis, with no bounding box or segmentation shown to the annotator. An annotator was instructed to label the user's left hand, the user's right hand, and other people's hands separately.

\subsection{System evaluation}
The interaction detection was evaluated using Leave-One-Subject-Out cross-validation. We applied our system to 9 participants from our dataset (participants \# 1-9), for whom interactions had been manually labelled and whose data had not been used to train the hand detection algorithms. In the cross-validation process, each participant in turn was left out for testing while the other 8 participants were used for training. Depending on the participant being left out, the training set of 8 participants on average consisted of $28,057\pm2,334$ frames ($935\pm77$ seconds) of interaction (48\%) and $30,850\pm3369$ frames ($1,028\pm112$ seconds) of no interaction (52\%). The participant in the testing set consisted on average of $3,507\pm3,369$ frames ($116\pm112$ seconds) of interaction (48\%) and $3,856\pm2,334$ frames ($128\pm77$ seconds) of no interaction (52\%). The classification was compared with manually labelled data.

\begin{figure*}
	\centering
	\begin{subfigure}[b]{0.3\textwidth}
	\includegraphics[width=\textwidth]{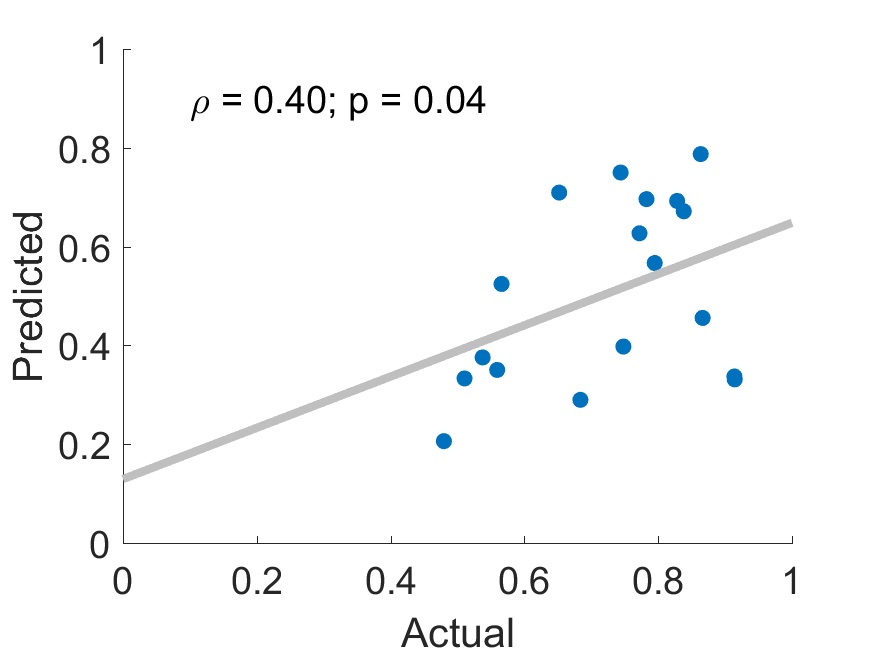}
	\captionsetup{justification=centering}
	\caption{}
	\label{fig:F3a}
	\end{subfigure}
	\begin{subfigure}[b]{0.3\textwidth}
	\includegraphics[width=\textwidth]{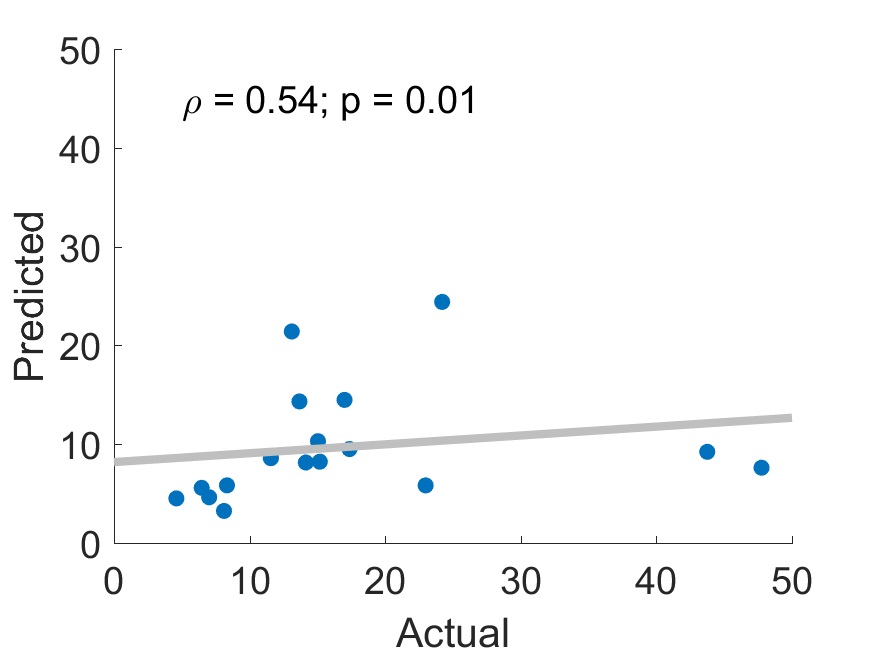}
	\captionsetup{justification=centering}
	\caption{}
	\label{fig:F3b}
	\end{subfigure}
	\begin{subfigure}[b]{0.3\textwidth}
	\includegraphics[width=\textwidth]{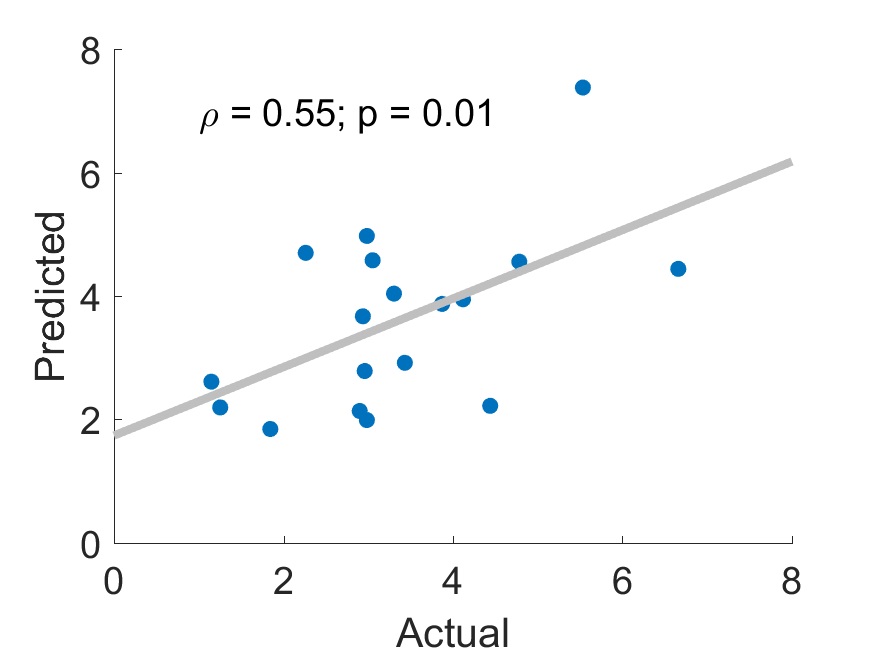}
	\captionsetup{justification=centering}
	\caption{}
	\label{fig:F3c}
	\end{subfigure}
\caption{Hand use metrics. Scatter plots comparing the interaction metrics predicted from the algorithm (y-axis) with the actual value from the human observer (x-axis), for each of the three proposed metrics in both hands (left and right hand). (a) Proportion of interaction over total recording time, (b) average duration of interactions (seconds), and (c) number of interactions per hour. The result of a Pearson correlation is shown for (a) and (c) because the data were normally distributed, while (b) was calculated with a Spearman correlation}
\label{fig:F3}
\end{figure*}

\subsection{Extraction of functional measures}

In order to translate the frame-by-frame results into more easily interpretable measures, the binary output of hand-object interaction detection was processed to extract: the amount of total interaction as a percentage of testing time, average duration of individual interactions in seconds, and number of interactions per hour (i.e., number of interactions normalized to video segment duration). 

Hand-object interaction outputs from the algorithm were assigned to one of two timelines, depending on arm angle: user's left hand and user's right hand. Failures in hand detection or segmentation could result in missing frames at the interaction detection stage. To address this issue, interactions were prolonged for 90 frames (3 seconds) if the hand was suddenly lost. Outside of this range, the frame was classified as a non-interaction. Furthermore, a moving average filter was applied to the binary frame-by-frame timelines of interaction. The moving average promotes temporal smoothness in the output, reduces the impact of labelling errors in the start and stop of interaction, and corrects minor faulty hand detection and segmentation. The filter window was chosen to have a length of 120 frames (corresponding to 4 seconds), with equally weighted samples. This duration was optimized empirically on the basis of its ability to meaningfully summarize the number and duration of underlying activities. The moving average was similarly applied to the manually labelled timelines. 

The output of the moving average was then normalized by subtracting the minimum value over the entire video and dividing by the difference of maximum and minimum values. The result was thresholded at 0.5 (i.e. values greater than 0.5 were considered to be interactions). Number and duration of interactions were extracted from this filtered output.

For each of the three metrics, the correlation between the algorithm output and the manually labeled data over the 9 tested participants was computed. The resulting correlations were tested against a hypothesis of no correlation using a one-tailed (right) test. Significance was set at $\rho=0.05$. Where data were normally distributed, a Pearson Correlation was chosen, otherwise a Spearman correlation was used.

\section{Results}

\subsection{Training and testing sets}

The demographic and injury characteristics of the participants are provided in Table \ref{T1}. The interaction detection was evaluated on participants \# 1–9. This testing set consisted of 8 males and 1 female with an average age of $52\pm13$ years and an average upper extremity motor score (UEMS) of $17\pm4$. As mentioned previously, participants \# 10–17 were used for the training of the preprocessing steps, namely the CNN-based hand detection and the classifier based on the Haar-like arm angle features. This hand detection training set consisted of 7 males and 1 female with an average age of $47\pm11$ years and an average UEMS of $20\pm3$.

This separation of the dataset was designed such that the final interaction detection evaluation was conducted on participants (\# 1–9) whose data had not been used in training any parts of the preprocessing algorithms. The two subsets have similar characteristics in terms of gender, age and UEMS.

\subsection{Interaction detection}

The filtered frame-by-frame interaction detection results are quantified using classification accuracy and F1–score (harmonic mean of precision and recall) for each participant; these are shown in Table \ref{table:T2}.  The average accuracy and F1–score are calculated for all the participants in the interaction detection evaluation. Over the 9 participants, the F1–scores were $0.74\pm0.15$ for the left hand and $0.73\pm0.15$ for the right hand. The accuracies were $0.70\pm0.16$ for the left hand and $0.68\pm0.18$ for the right hand.

\subsection{Functional hand use}
The three proposed interaction metrics were compared between the automated output and the manual observations, and the results are shown in Fig. \ref{fig:F3}. For all three metrics (proportion of interaction time, average duration of interactions, and number of interactions per hour), a moderate, significant correlation was found between the actual and predicted metrics ($\rho=0.40$, $p=0.04$; $\rho=0.54$, $p=0.01$;$ \rho=0.55$, $p=0.01$, respectively).

\subsection{Features analysis}

Moreover, we sought to understand the importance of each feature for hand-object interaction. This involved selecting only one type of feature at a time (optical flow, HOG and colour histogram) in the classifier. The overall average accuracy and F1-score for all subjects are articulated in Table \ref{T2}. There was a general trend towards relatively slight decreases in performance when features were removed, with the resulting F1-scores ranging from $0.66\pm0.17$ to $0.73\pm0.14$, and accuracies ranging from $0.66\pm0.15$ to $0.69\pm0.12$.

\begin{figure*}
	\centering
	\begin{subfigure}[b]{0.72\textwidth}
	\textbf{(a)}
	\includegraphics[width=\textwidth]{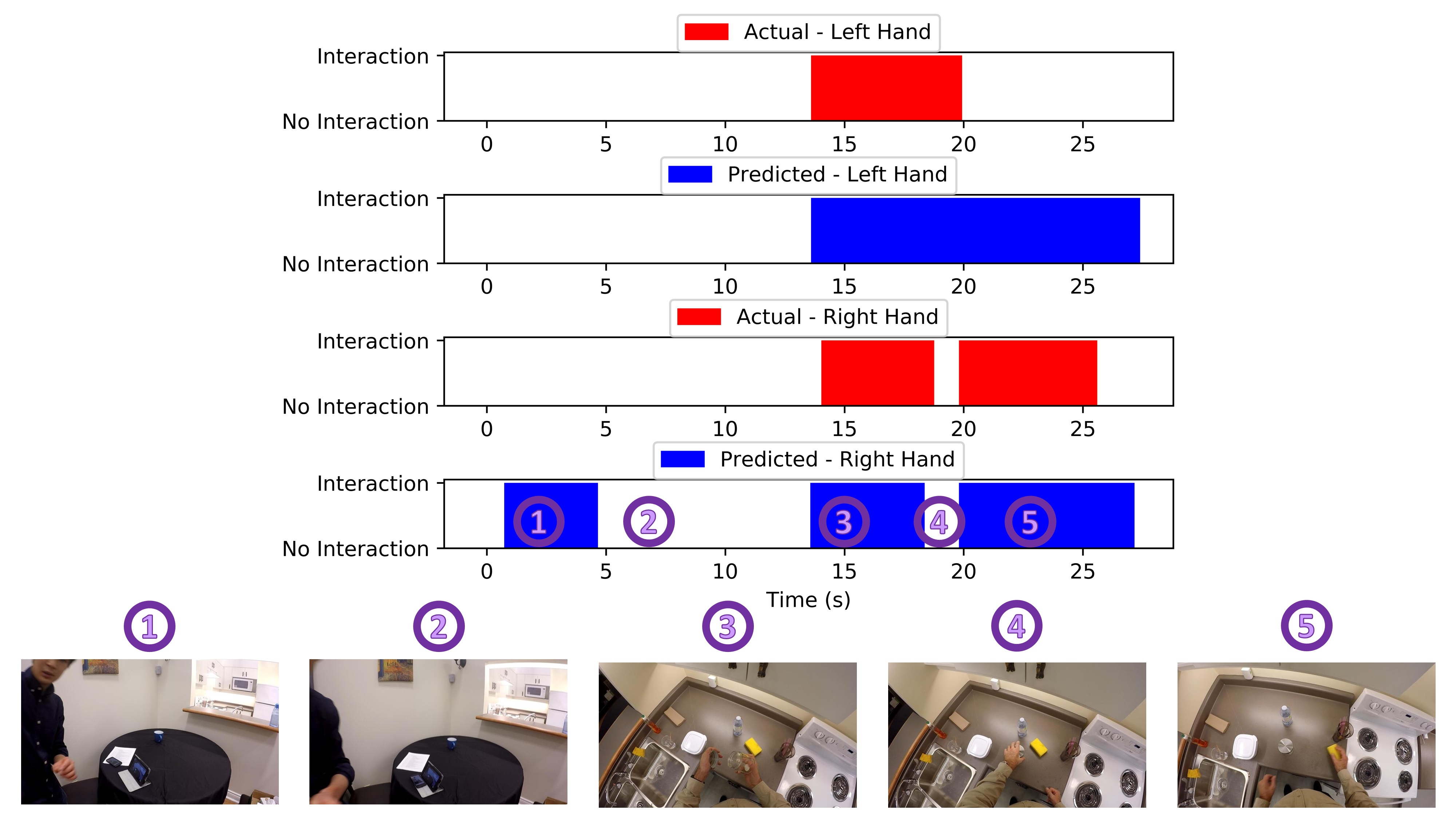}
	\label{fig:F4a}
	\end{subfigure}
	\begin{subfigure}[b]{0.72\textwidth}
	\textbf{(b)}
	\includegraphics[width=\textwidth]{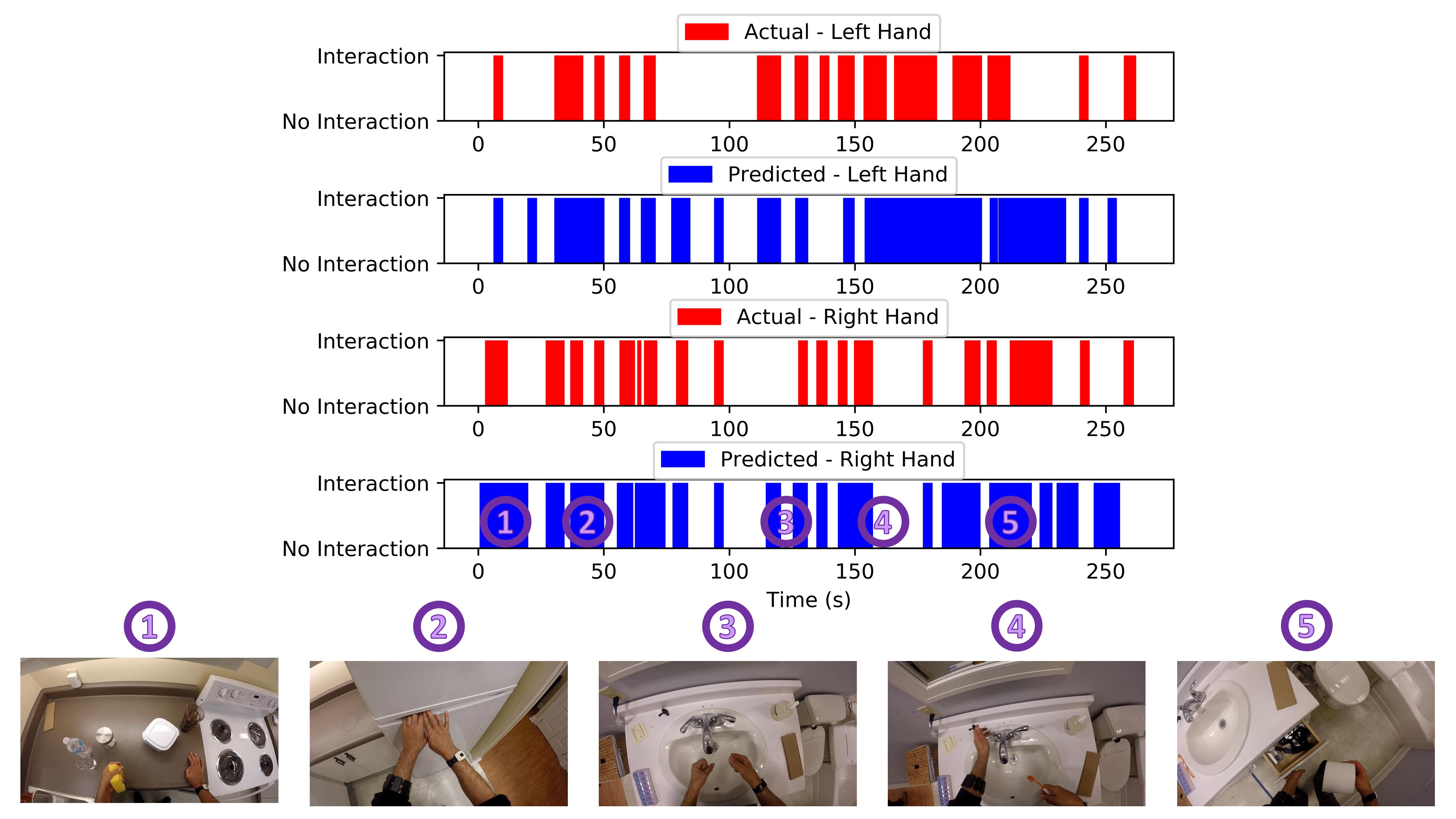}
	\label{fig:F4b}
	\end{subfigure}
	\begin{subfigure}[b]{0.72\textwidth}
	\textbf{(c)}
	\includegraphics[width=\textwidth]{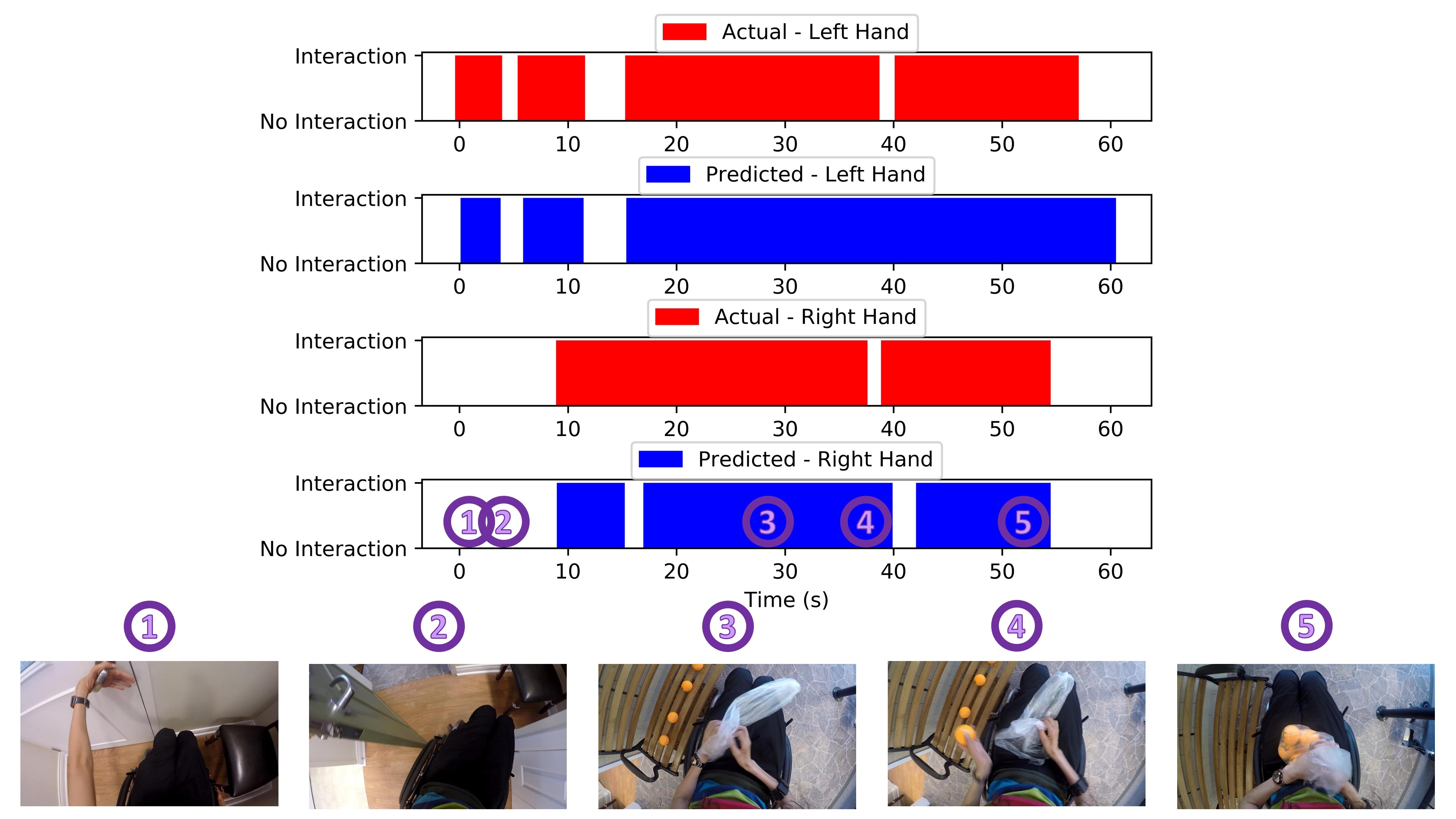}
	\label{fig:F4c}
	\end{subfigure}
\caption{Example binary hand-object interaction graphs of 3 participants. The graphs compare the predicted interactions from the algorithm output to the actual interactions from the manually labeled data, after applying the moving average filter. Example frames of the activities in different segments of videos are shown underneath. a) Participant \# 2. b) Participant \# 5. c) Participant \# 9. Note that in some cases the videos were briefly paused in between the activities shown} 
\label{fig:F4}
\end{figure*}

\section{Discussion}

Outcome measures capable of capturing the hand function of individuals with SCI in their home environments are needed to better assess the impact of new interventions. Current methods are limited to self-report or direct observation. In this study, we demonstrated the feasibility of measuring the functional hand use of individuals with UE dysfunction resulting from cervical SCI using a novel wearable system. The system is based on a wearable camera and a custom algorithmic framework capable of automatically analyzing hand use in ADLs at home. The frequency and duration of hand-object interaction could potentially serve as the basis for new outcome measures that provide clinicians and researchers with an objective measure of an individual’s independence in UE tasks.

We demonstrated the criterion validity of an interaction-detection system by comparing the algorithm outputs to manually labelled data, taken here to be the gold standard. The binary hand-object interaction classification was shown to be robust across multiple ADLs, environments, and individuals, with an average F1-score of $0.74\pm0.15$ for the left hand and  $0.73\pm0.15$ for the right hand after the moving average (Table \ref{table:T2}). For the system evaluation, we sampled activities at random among participants; thus, each participant had a unique set of activities during the leave-one-subject-out cross-validation, meaning that the system was tested on ADLs that it had not necessarily seen before. The evaluation was performed in a non-scripted manner; however, the tasks were specified (i.e. participants were only given the name of the tasks with no instruction on how these should be performed). The dataset generated was balanced between the proportion of interaction and non-interaction frames (48 and 52\% respectively).

As visible in Table \ref{T2}, performance varied between participants and sometimes between hands. It is important to note that since subsets of the data were labelled and used for the interaction detection evaluation, participants were not all evaluated on the same activities. This difference in activities may account for variations in F1-score and accuracy across individuals. For example, in participant \# 1, the data used consisted largely of activities in the living room and involved finer manipulation activities such as eating potato chips and picking up a coin. However, perhaps the main factor that resulted in faulty hand detections was scenes consisting of wooden objects and flooring. Based on qualitative observation the R-CNN tended to produce false positives on wood. While the Haar-like methods are able to eliminate the majority of the false positives, their impact is greater in certain environments and scenes than others. The inconsistency observed is also related to differences in classifier performance between the left and right hand in some cases, for example in participant \# 4. Direct observation of the frames again point to the poor detection of the right hand, which is closer to the wooden floor, leading to some missed identifications. It is therefore a point of limitation given that the hand-object interaction directly relies on this pre-processing step due to the serial architecture of the pipeline. We anticipate that improvements in hand detection will improve the performance of the overall pipeline. Although minor, variations in task and injury characteristics may also contribute to lower accuracy. For example, participant \# 6 used a joystick to control their wheel chair. Such activities involved a relatively small movement, which can be difficult to reliably identify in the egocentric video. Aspects of participant postures post-injury, such as pronounced curving of the back, were also found to reduce the Field of View (FOV) of the camera, resulting in the hand not appearing in the frames.

The functional measures of interest are the number and duration of hand-object interactions. Given the variability in activities and labelled video durations between participants, the measures were summarized using the amount of total interaction normalized to testing time, the average duration of individual interactions, and the number of interactions per hour. A moving average was applied to reduce noise from short-term fluctuations caused by a potentially faulty preprocessing step (e.g. hand detection or segmentation), misclassification, or fast interaction sub-tasks within an activity. Perhaps most important is that the moving average highlights important long-term trends that are obscured by a framewise analysis of long recordings. Consider the timeline in Fig. \ref{fig:F4}, where the moving average of the binary interaction graph is plotted against time. Here, the transitions between tasks are clearly observed as changes between the interaction and non-interaction classes. This example illustrates the value of capturing these metrics for the measurement of hand use at home.

The functional measures of interest are the number and duration of hand-object interactions. Given the variability in activities and labelled video durations between participants, the measures were summarized using the amount of total interaction normalized to testing time, the average duration of individual interactions, and the number of interactions per hour. A moving average was applied to reduce noise from short-term fluctuations caused by a potentially faulty preprocessing step (e.g. hand detection or segmentation), misclassification, or fast interaction sub-tasks within an activity. Perhaps most important is that the moving average highlights important long-term trends that are obscured by a framewise analysis of long recordings. Consider the timeline in Fig. \ref{fig:F4}, where the moving average of the binary interaction graph is plotted against time. Here, the transitions between tasks are clearly observed as changes between the interaction and non-interaction classes. This example illustrates the value of capturing these metrics for the measurement of hand use at home.

The comparison of the predicted and actual values for the interactions per hour, average duration of interactions, and proportion of interaction time for every participant are summarized in Fig. \ref{fig:F3},. These results indicate to what extent the conclusions about hand use provided by the automated system correlate to the results from a human observer. We found moderate and significant correlations, which provides evidence for the viability of this approach, while highlighting the need for continued improvements in the algorithms used for interaction detection. Deviations between the predicted and actual metrics are likely a combination of errors in hand detection, interaction detection, and the filtering process, providing multiple targets to increase performance. It is also possible that outliers in the limited number of participants in this evaluation may have partially skewed the results, and that a larger dataset would yield a more robust outcome (e.g., consider the influence of the two data points with high actual values but low predicted values in Fig. \ref{fig:F3b}).

\begin{table}
\small\sf\centering
\caption{F1-Score and accuracy for left (L) and right (R) hand for each participant as well as the average for each of the feature
\label{T2}}
\label{table:T2}
\begin{tabular}{ccccc}
\toprule
\texttt{} & \multicolumn{2}{l}{F1-score} & \multicolumn{2}{l}{Accuracy} \\
\texttt{} & \texttt{L} & \texttt{R} & \texttt{L} & \texttt{R} \\
\midrule
Participant \# & \multicolumn{2}{c}{} & \multicolumn{2}{c}{} \\
\midrule
\texttt{1} & \texttt{0.54} & \texttt{0.53} & \texttt{0.42} & \texttt{0.42} \\
\texttt{2} & \texttt{0.60} & \texttt{0.75} & \texttt{0.73} & \texttt{0.79} \\
\texttt{3} & \texttt{0.86} & \texttt{0.73} & \texttt{0.79} & \texttt{0.63} \\
\texttt{4} & \texttt{0.85} & \texttt{0.58} & \texttt{0.79} & \texttt{0.59} \\
\texttt{5} & \texttt{0.72} &  \texttt{0.67} & \texttt{0.75} & \texttt{0.70} \\
\texttt{6} & \texttt{0.55} & \texttt{1.00} & \texttt{0.48} & \texttt{0.99} \\
\texttt{7} & \texttt{0.84} & \texttt{0.80} & \texttt{0.82} & \texttt{0.69} \\
\texttt{8} & \texttt{0.78} & \texttt{0.63} & \texttt{0.67} & \texttt{0.51} \\
\texttt{9} & \texttt{0.93} & \texttt{0.91} & \texttt{0.88} & \texttt{0.88} \\
\makecell[c]{Mean \\ $\pm$S.D.} &  \makecell[l]{$0.74$ \\ $\pm0.15$} & \makecell[l]{$0.73$ \\ $\pm0.15$} & \makecell[l]{$0.70$ \\ $\pm0.16$} & \makecell[l]{$0.68$ \\ $\pm0.18$} \\
\midrule
Feature & \multicolumn{2}{l}{} & \multicolumn{2}{l}{}\\
\midrule
Optical Flow & \makecell{$0.73$ \\ $\pm0.14$} & \makecell{$0.70$ \\ $\pm0.13$} & \makecell{$0.68$ \\ $\pm0.16$} & \makecell{$0.66$\\$\pm0.15$} \\
HOG & \makecell{$0.72$ \\ $\pm0.12$} & \makecell{$0.72$\\$\pm0.14$} & \makecell{$0.69$\\$\pm0.12$} &\makecell{$0.68$\\$\pm0.15$} \\
Colour Histogram  & \makecell{$0.70$\\$\pm0.12$} & \makecell{$0.66$ \\ $\pm0.17$} & \makecell{$0.68$ \\ $\pm0.10$} & \makecell{$0.66$ \\$\pm0.16$} \\
\bottomrule
\end{tabular}
\end{table}

A challenge in extracting hand use metrics based on frame-by-frame interaction detections is that complex timelines of interactions (Fig. \ref{fig:F4}) need to be simplified into simple outcome units or scores. Further exploration is warranted in how to best summarize the interactions taking place over time among a variety of activities, and in optimizing the normalization used for comparing different recording times.
 
To the best of our knowledge, the hand-object interaction problem has not been studied before in individuals with neurological injuries. Thus, we sought to understand what types of information were most beneficial to accurate interaction detection. This understanding is crucial in that hand movements or postures may vary substantially between individuals with SCI having different patterns of injury or impairment. The analysis in Table \ref{table:T2} revealed that all the features (optical flow, HOG and colour histogram) were able to contribute to the hand-object interaction classification. The combined features produced the overall highest performance, but the gain over using a single feature type was minimal. The average F1-score when using only optical flow was the highest, followed by HOG, and then closely by colour histogram. In contrast, in our previous work with able-bodied participants \cite{RefWorks:170}, using only HOG was found to be more useful than using only optical flow. The fact that this finding no longer held true with participants with SCI suggests that relying on shape (i.e., hand posture) may be less beneficial in the presence of varying levels of impairment and compensatory postures. Classifiers tailored to different types and severities of injury could potentially increase the performance achievable with shape features. 

Our validity evaluation focused on a comparison with manually labelled interaction data, and we did not attempt to correlate the results with existing outcome measures. The collected dataset consisted of standardized and pre-determined activities. As such, it is not representative of the frequency of hand use in a natural home environment and includes tasks that some participants may not normally perform on their own. While we expect that individuals with better hand function may independently perform UE tasks more frequently in their daily life, evaluating this relationship will require collecting data in the home in an unscripted manner. In the present study, UEMS data was collected for demographic and study inclusion purposes, but its relationship to the interaction metrics was not investigated, for the reasons above. With appropriate data collection in the home, an important extension of our analysis will include the comparison of the interaction-detection metrics with other UE outcomes measures. For example, greater independent use of the hand would be expected to correlate with the Graded Redefined Assessment of Strength, Sensibility and Prehension (GRASSP), which is sensitive to fine gradations in hand function after SCI, or the Spinal Cord Independence Measure III (SCIM), which aims to quantify independence in the community. Of note, the actual amount of hand use at home (corresponding to Performance in the revised ICF model \cite{RefWorks:139}) represents a different construct than hand impairment or the capacity to perform activities, which are more closely captured by the GRASSP and SCIM. Nonetheless, some degree of relationship is still expected. While these investigations will be the subject of future work in parallel with continued algorithm improvements, the present study was required to validate the feasibility of the interaction detection step before metrics derived from it (i.e., number and duration of interactions) can be meaningfully compared to other measures.

Another limitation in this study is that the hands of individuals other than the user were not analysed. At home, it would be of great interest to quantify caregiver assistance by tracking their manipulations of objects in front of the individual with SCI. The system described supports the detection of caregiver hands and their associated interactions (a valuable metric for reliance on care). However, limited data of this type was collected in this study, because participants most often chose to complete the tasks on their own or skip them completely, instead of asking for assistance. This behaviour can likely be explained by the lack of familiarity with the researchers and the pressure to perform in a research setting. In the home setting, however, we expect that more caregiver actions will be captured by the system, which will provide important information about independence. 

There are remaining technical challenges to be explored beyond this study, namely the improvement of the preprocessing steps and computation time. The detection and segmentation of the hand are based on skin and hand shape characteristics, which can be influenced by glare and objects with a similar shape or colour to the hand (e.g. a wooden floor, table, or door). The performance of the overall hand-object interaction system is dependent on the accuracy of the hand detection and segmentation steps, and can be expected to improve with them. For example, in a preliminary analysis, only manually selected frames with good hand segmentation were used, and the interaction detection performance was found to be 0.81 for F1-score in 15,471 frames from 3 participants \cite{NationalSCI2017}. Secondly, to avoid privacy and usability concerns for the user \cite{Likitlersuang2017}, the ideal system should be mobile and process the video in real-time, storing only the extracted metrics and not the video. Unfortunately, the proposed algorithm remains computationally expensive for a mobile system. Its take on average 2.70 seconds per frame to process hand-object interaction from the input image frames to the output metrics (Intel-i7-8700k\@4.8GHzOC, DDR4-16GB\@3200MHzOC, GTX1080TiOC-GDDR5X-11GB, Ubuntu14.04 LTS 64-bit). Lastly, the performance of the algorithms will need to be evaluated in a wider range of environments, with challenges that may include imperfect lighting, differences in camera orientation, and more diverse tasks. Improvements using recent computer vision techniques for hand tracking and segmentations \cite{Redmon2017}, as well as better feature selection, have the potential to improve performance and speed. 

\section{Conclusions}

In this study, we demonstrated the potential of an egocentric wearable camera system for capturing an individual’s functional hand use in the home environment. Novel outcome measures based on this system have the potential to fill the research gaps in home-based assessment of the UE in neurorehabilitation, which currently relies heavily on self-report. We have demonstrated the feasibility of the interaction-detection process and illustrated how this concept can be used to derive meaningful outcome measures, such as the number and duration of independent, functional object manipulations. More broadly, our study provides a framework for future research in UE assessment within the broader community.

\section{Availability of data and materials}

The dataset will be shared for academic purposes upon reasonable request. Please contact author for data requests.

\section{Abbreviations}
\textbf{ADLs:} Activities of daily living
\\
\textbf{ANS SCI:} Adaptive Neurorehabilitation Systems Laboratory dataset of participants with spinal cord injury
\\
\textbf{AOTA:} American Occupational Therapy Association
\\
\textbf{CNN:} Convolutional neural network
\\
\textbf{cSCI:} Cervical spinal cord injury
\\
\textbf{faster R-CNN:} Faster regional convolutional neural network
\\
\textbf{GRASSP:} Graded Redefined Assessment of Strength, Sensibility and Prehension
\\
\textbf{HOG:} Histograms of oriented gradients
\\
\textbf{ICF:} International classification of functioning, disability and health
\\
\textbf{PCA:} Principal component analysis
\\
\textbf{SCI:} Spinal cord injury
\\
\textbf{SCIM:} Spinal Cord Independence Measure III
\\
\textbf{UE:} Upper extremity

\section{Acknowledgments}
The authors would like to thank Pirashanth Theventhiran for his valuable assistance in the data labelling process. The authors would also like to thank all the participants of the study.

\subsection{funding}
This study was supported in part by the Natural Sciences and Engineering Research Council of Canada (RGPIN-2014-05498), the Rick Hansen Institute (G2015–30), and the Ontario Early Researcher Award (ER16–12-013).

\section{Acknowledgments}

\subsection{Affiliations}
Institute of Biomaterials \& Biomedical Engineering, University of Toronto, Toronto, Ontario, Canada\\ \\
Jirapat Likitlersuang, Elizabeth R. Sumitro, Ryan J. Visée \& José Zariffa
KITE, Toronto Rehabilitation Institute, University Health Network, Toronto, Ontario, Canada
Jirapat Likitlersuang, Elizabeth R. Sumitro, Tianshi Cao, Ryan J. Visée, Sukhvinder Kalsi-Ryan \& José Zariffa\\ \\
Department of Physical Therapy, University of Toronto, Toronto, Ontario, Canada
Sukhvinder Kalsi-Ryan 

\subsection{Contributions}
JL, SKR and JZ designed the study. JL and ERS carried out the data collection. JL, TC, and RJV participated in algorithm development and implementation. JL and JZ drafted the manuscript. All authors read and approved the final manuscript.

Corresponding author
Correspondence to José Zariffa.

\bibliography{referenceV1}
\bibliographystyle{unsrt}

\end{document}